\documentclass[aps,prl,twocolumn,showpacs,superscriptaddress,groupedaddress]{revtex4}  
\usepackage{graphicx}  
\usepackage{dcolumn}   

\usepackage{psfrag}
\usepackage[T1]{fontenc}
\usepackage{epsfig}
\usepackage{times}
\usepackage{amsmath}
\usepackage{amssymb}
\usepackage{graphicx}
\usepackage{wasysym}
\usepackage{textcomp}

\usepackage{bm}        
\usepackage{amssymb}   
\newcommand{\be}{\begin{equation}}
\newcommand{\ee}{\end{equation}}

\begin{document}



\title{Classical and Quantum Annealing in the Median of Three Satisfiability}

\author{T. Neuhaus}
\email{t.neuhaus@fz-juelich.de}
\affiliation{%
Institute for Advanced Simulation,
J\"ulich Supercomputing Centre,
Forschungszentrum J\"ulich,
D-52425 J\"ulich, Germany
}%
\affiliation{%
Fakultät für Physik, Universit\"at Bielefeld, D-33501 Bielefeld, Germany
}%
\author{M. Peschina}
\email{ma.peschina@fz-juelich.de}
\affiliation{%
Institute for Advanced Simulation,
J\"ulich Supercomputing Centre,
Forschungszentrum J\"ulich,
D-52425 J\"ulich, Germany
}%
\affiliation{%
Institut f\"ur Kernphysik, Johannes Gutenberg-Universit\"at Mainz, D- 55128 Mainz, Germany
}%
\author{K. Michielsen}
\email{k.michielsen@fz-juelich.de}
\affiliation{%
Institute for Advanced Simulation,
J\"ulich Supercomputing Centre,
Forschungszentrum J\"ulich,
D-52425 J\"ulich, Germany
}%
\author{H. De Raedt}
\email{h.a.de.raedt@rug.nl}
\affiliation{%
Department of Applied Physics,
Zernike Institute of Advanced Materials,
University of Groningen, Nijenborgh 4, NL-9747 AG Groningen, The Netherlands
}%

\date{\today}
\begin{abstract}
We determine the classical and quantum complexities
of a specific ensemble of three-satisfiability problems
with a unique satisfying assignment
for up to $N=100$ and $N=80$ variables, respectively.
In the classical limit we employ generalized ensemble techniques and
measure the time that a Markovian Monte Carlo process spends in
searching classical ground states. In the quantum limit we determine
the maximum finite correlation length
along a quantum adiabatic trajectory determined by the linear sweep of the adiabatic control parameter in
the Hamiltonian composed of the problem Hamiltonian and the constant transverse field Hamiltonian.
In the median of our ensemble
both complexities diverge exponentially
with the number of variables. Hence, standard, conventional adiabatic quantum computation
fails to reduce the computational complexity to polynomial.
Moreover, the growth-rate constant in the quantum limit is 3.8 times
as large as the one in the classical limit, making classical fluctuations
more beneficial than quantum fluctuations in ground-sate searches.
\end{abstract}


\maketitle

At the borderline of mathematics and physics lie optimization problems, that
can be cast into solving a minimization problem on a discrete set of variables:
Given a scalar cost function $H_0(s)$ that is bounded from below by zero, and given a set of
integer variables (Ising spins) $s_i=\pm 1$ with $i=1,...,N$,
one may ask: Which assignment or - satisfying assignment - solves $H_0=0$ ?
Many satisfying assignment problems with Boolean variables $b_i=(1 +s_i)/2$ that are NP-hard
have precisely this form.
In this work we study the 3-satisfiability (3-SAT) problem, a NP-hard problem
at the heart of complexity theory \cite{Cook},
by means of methods used in (quantum) statistical physics.


Under the assumption ${\rm P} \ne {\rm NP}$,
the computational effort for any classical algorithm
to solve NP-hard problems is believed to be
${\cal O} (e^{g N})$, where
$g$ denotes the growth-rate constant.
For a trivial classical or unstructured search, that is the evaluation of
$H_0$ over all configurations, $g = \ln 2 $.
In quantum computation this search finds its analog in Grover's algorithm \cite{Grover}
with $g = \ln 2/2 $.
A polynomial solution to a NP-hard problem is expected to have $g=0$.


Conventional, standard adiabatic quantum computation (AQC) \cite{Nishimori,Farhi} assumes a linear interpolation between
the NP-hard problem Hamiltonian $H_0$ and a non-commuting driver Hamiltonian
$H_D = \sum_i \sigma^x_i$ (the ``transverse field''), where $\sigma^x_i$ is the $x$-component of the Pauli matrix.
A statistical analysis of AQCs determines the thermodynamic and quantum singularities of
the partition function
\be
Z_{\rm AQC}(\beta,\lambda) = {\rm Tr}~
e^{-\beta\{(1-\lambda) H_D + \lambda H_0\}},
\label{partition_function}
\ee
where $\beta=1 / k_B T$ denotes the inverse temperature, $k_B$
Boltzmann's constant and $0 \le \lambda \le 1$ the quantum adiabatic
control parameter.
In the vicinity of the point $P_{\rm 0}^*=(\beta,\lambda)=(\infty,1)$ the optimization
problem is solved, as vanishing thermal and quantum fluctuations lead to
the exact ground-state.
We study the approach to $P_{\rm 0}^*$ from regions of large thermal, as well as large
quantum fluctuations on lines of parameters $\beta$ and $\lambda$. In particular, we study
measures of complexity in the classical limit at $\lambda=1$ as a function
of $\beta$ and in the pure quantum limit at almost zero temperature as a function of $\lambda$.

In the classical limit, a measure of complexity is the Monte Carlo (MC)
search time for the ground state in multicanonical ensemble and Wang-Landau simulations \cite{Muca,WangLandau}.
These MC simulations perform a Markovian process with random walk dynamics in the energy.
We count the number of MC steps in-between ground-state findings in the mean.

In the pure quantum limit we determine the maximal spin-spin correlation length
$\xi_{\rm max}$, i.e. the inverse of the first energy gap at the presumed quantum
phase transition at $\lambda^*$, from the exponential decay of a two point function in imaginary time
(see Eq.~(\ref{quantum_correlator}).
If there exists an avoided level crossing, the spin-spin correlation length is
finite for a finite number of spins.
In accordance with Landau Zener theory \cite{LandauZener},
in AQCs the running time of ground-state searches is limited to a
time scale ${\cal T}$ of order ${\cal O} (\xi^{2}_{\rm max})$
from below.
Hence, for a NP-hard problem, a spin-spin correlation length growing
exponentially with $N$ would yield a computational complexity for quantum ground-state searches
that is similar to the one expected for a classical search, and therefore would make AQC fail.

It is argued that exponentially small energy gaps can be induced
by the presence of first-order phase transitions, hampering the performance of AQC
in optimization problems related to the 3-SAT problem~\cite{Schuetzhold,Amin_2009,Jorg_2010,Altshuler_2010}.
Their predictive power
for specific optimization problems
is however limited, as
any first order phase
transition may - or just may not - turn into
second order at a critical point.
A particular nasty situation is
encountered, if there is either a weak
first - or second order phase transition, a situation
that has recently been studied
in the exact cover problem~\cite{Young_both}.

The 3-SAT problem is defined
on a set of $i=1,...,N$
classical Ising spins $s_i = \pm 1$.
Its Hamiltonian can be written as a sum of $M$ three-point functions, called clauses:
%
\be
H_0= \sum_{\alpha=1}^M \Upsilon^{3}
(
   \epsilon_{\alpha,1} s_{i[\alpha,1]}
  ,\epsilon_{\alpha,2} s_{i[\alpha,2]}
  ,\epsilon_{\alpha,3} s_{i[\alpha,3]}
).
\ee
The function $\Upsilon^{3}$ results from a transcription of the disjunctive cubic Boolean form of a clause $b_l \lor b_k \lor b_m$
to Ising degrees of freedom:
\begin{eqnarray}
\Upsilon^{3}(s_k,s_l,s_m)= \frac{1}{8}
\lbrace
  ( s_k s_l s_m )
+ ( s_k s_l + s_l s_m + s_k s_m ) \nonumber \\
+ ( s_k + s_l + s_m )
- 1
\rbrace ,
\end{eqnarray}
%
with $k,l,m\in\{1,\ldots,N\}$ and $k\neq l\neq m$.
If $p=1,2,3$ denotes the position of a spin
within clause $\alpha$ and if $\eta$ labels
an element from a set of $N_{\eta}$
realizations of spins $s_i$,
the quantity  $\epsilon_{\alpha,p}^{\eta}$
takes values $\pm 1$.
The symbol $i[\alpha,p]$ denotes a map
$ [\alpha , p ] \rightarrow i $
from indices of the clause to the set of spins.

\begin{figure}[t]
\centering
\psfrag{LABEL1}[t][t][2][0]{$\beta$}
\psfrag{LABEL2}[t][t][2][0]{$\langle H_0\rangle$}
\psfrag{LABEL3}[t][t][2][0]{$\beta_{\rm CMAX}$}
\psfrag{LABEL4}[t][t][2][0]{$o_{\rm gs}$}
\psfrag{LABEL5}[t][t][2][0]{${\rm ln}~{P}(o_{\rm gs})$}
\psfrag{LABEL6}[t][t][2][0]{$P_{\rm max, left}$}
\psfrag{LABEL7}[t][t][2][0]{$P_{\rm min}$}
\includegraphics[angle=-90,width=0.55\textwidth]{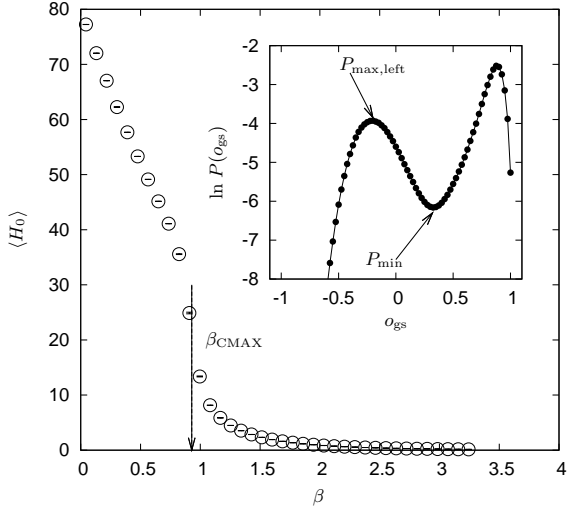}
\caption{Classical limit: Expectation value of the problem Hamiltonian
$\langle H_0\rangle$ as a function of $\beta$ for a particular realization
with $N=80$ and $r=8$. Arrow: Position of the maximal specific heat.
Inset: Distribution function
of the ground-state overlap observable, see Eq.~(\ref{overlap_order_parameter_distribution}).
$P_{\rm max, left}$ and $P_{\rm min}$ determine
the nucleation barrier $B_0$.
\label{fig:fig1}}
\end{figure}

The 3-SAT problem has been studied on
{\it random} instances with statistical methods, the hardness
being characterized by the clauses-to-variables ratio $r=M/N$.
It exhibits at least two phases: a ``SAT'' phase
for $r < 4.2 $ with degenerate ground-states and an ``UN-SAT'' phase
for $r > 4.2$ where satisfying assignments
are exponentially rare \cite{Kirkpatrick_1999}.
Note that also computationally demanding instances can be constructed for $r=3$ \cite{Znidaric_2005}.

We study instances with unique satisfying assignments (USA) and for which $r$ is largely free.
For this purpose we generate random ground-states
${s_1^{\rm gs},...,s_N^{\rm gs}}$
and solve $\epsilon_{\alpha,1} s_{i[\alpha,1]}^{\rm gs}=1$ for a particular
map $i[\alpha,1]={\rm mod}(\alpha-1,N)+1$ and for $\alpha=1,...,M$.
The remaining $\epsilon_{\alpha,p}$ and $i[\alpha,p]$ for $p=2,3$
are generated at random with the help of MC updates.
We use various filter techniques to exclude non-USA instances.

In studies of the classical limit for $r=5,8$ we
use multicanonical ensemble and Wang-Landau simulations \cite{WangLandau,Muca}
and determine a statistical estimate of
the density of states $n(E)$ on the entire
discrete energy interval $0 \le E \le E_{\rm max}$.
We obtain the canonical partition function
$Z_{\rm can}(\beta)$ via its spectral representation
and calculate thermodynamic quantities like the internal
energy $\partial_\beta {\rm ln} Z_{\rm can}(\beta)$
and the specific heat  $\beta^{-2}\partial^2_\beta {\rm ln} Z_{\rm can}(\beta)$.
Numerical analysis shows a discontinuous phase transition at some value $\beta^*$, see Fig.~\ref{fig:fig1}.
Its first order nature
can best be established by considering the ground-state overlap observable
$o_{\rm gs}= N^{-1}\sum_{i=1}^N s_i s_i^{\rm gs}$
and its distribution function at the
specific heat peak position $\beta_{\rm CMAX} \approx \beta^*$
\be
P(o_{\rm gs}) =  Z^{-1}\mathop{{\sum}'}
\delta\left[o_{\rm gs}-\frac{1}{N}\sum_{i=1}^N s_i s_i^{\rm gs}\right]~e^{-\beta_{\rm CMAX}E} ,
\label{overlap_order_parameter_distribution}
\ee
where $\sum^{\prime}$ denotes a sum over all spin configurations.
For almost all realizations $P(o_{\rm gs})$
exhibits a bimodal shape with one sharp peak
at $o_{\rm gs,right} \approx 1$ with a value $P_{\rm max,right}$,
an example being depicted in the inset of Fig.~\ref{fig:fig1}.
The second peak with value $P_{\rm max,left}$ is well separated
from the first one and is located at $o_{\rm gs,left}$.
At $r=8$ a finite size scaling analysis
yields a non-vanishing overlap order parameter gap
$\langle\Delta o_{\rm gs}\rangle=\langle o_{\rm gs,right}\rangle-\langle o_{\rm gs,left}\rangle=1.11(1)$ for
the thermodynamic limit and in the mean of realizations.
Discontinuous phase transitions are also associated with
nucleation free energy barriers. As the maxima in $P(o_{\rm gs})$ are separated by
a minimum with value $P_{\rm min }$, and in analogy to free
energy barrier definitions in Ising magnets and glasses,
we use Binder's method \cite{Binder_method} to define a nucleation barrier
$B_0= \ln [P_{\rm max,left } / P_{\rm min} ]:=B_0^\eta$ for
each realization $\eta$.

\begin{figure}[t]
\centering
\psfrag{LABEL1}[t][t][2][0]{$N$}
\psfrag{LABEL4}[t][t][2][0]{${\rm Complexities}$}
\psfrag{TITLETITLE1TITLE}[t][t][2][0]{$\ln \langle \tau_s^0/N^2\rangle$}
\psfrag{TITLETITLE2TITLE}[t][t][2][0]{$\langle B_0\rangle$}
\psfrag{LABEL3}[t][t][2][0]{${\rm ln}[{\tau_s^0/N^{2}}]$}
\psfrag{LABEL2}[t][t][2][0]{$B_0$}
\includegraphics[angle=-90,width=0.55\textwidth]{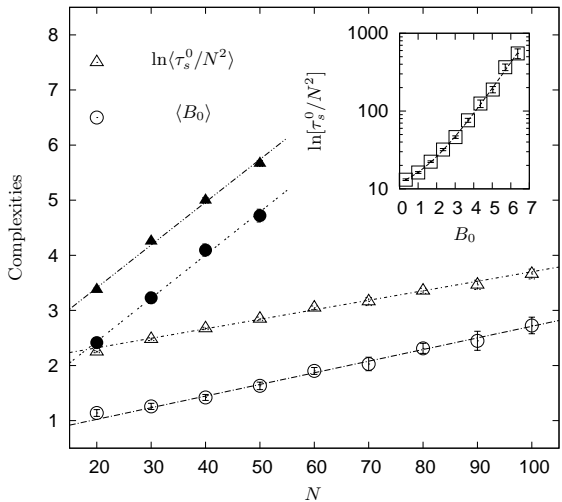}
\caption{Classical limit: Expectation values for the logarithmized search time
$\ln \langle\tau_s^0/N^{2}\rangle$ (triangles) and the nucleation barrier
$\langle B_0\rangle$ (circles) as a function of the number of spins
$N$. Straight lines correspond to exponential singularities.
Open symbols: $r=8$; full symbols: $r=5$.
Inset: $N=100$ and $r=8$ dependency
of ${\rm ln}[\tau_s^0/N^{2}]$ on $B_0$ within a set of 1000 realizations.
\label{fig:fig2}}
\end{figure}

We also perform MC simulations in the multicanonical partition function
$Z_{\rm muca} = \sum^{\prime} {\rm exp}[-\ln~n(E)]$
yielding an equal distribution
for the probability  $P_{\rm muca} (E)={\rm const}$ to find an energy
$E$ in the Markov chain.
The MC dynamics in the energy $E$
is different from that of a free random
walk with polynomial singularities
$\tau \propto N^2$ in autocorrelation times, as hidden free
energy barriers at energies in the vicinity of the ground-state $E=0$ slow
down the diffusion. Realization wise we measure the number of
local Metropolis update steps $\tau_s^0/N^2$ in the mean, that
an interacting walk in energy spends in the
"transition'' from $E_{\rm max}$ towards
the ground-state energy. The factor $N^{-2}$ corrects
for the trivial free walk behavior that is present even in the absence
of barriers.

Figure \ref{fig:fig2} displays the results for the averages of $\ln [\tau_s^0/N^2]$ (triangles) and $B_0$ (circles)
calculated in the median for $r=5,8$ and a number of spins up to $N=100$.
Both quantities exhibit exponential behavior
of the form ${\rm A}~{\rm exp}[g_c N]$
with classical growth-rate constants
$g_{c}^{\tau}=0.077(4)$ (barrier)
and  $g_c^{B}=0.078(3)$ (search time) for $r=5$.
For $r=8$ we find $g_c^{\tau}=0.016(1)$
and $g_c^{B}=0.021(2)$, respectively .
Hence, as expected, the complexity grows as
$r$ is lowered to $r^* \approx 4.2$ from above. Note however that
the growth-rates are much smaller than $g=\ln 2$
of an unstructured search.  The inset of Fig.~\ref{fig:fig2} displays
binned mean values of $\tau_s^0/N^2$ as a function of the
nucleation barrier $B_0$ for $r=8$ and $N=100$ for a set
of $N_\eta=1000$ realizations. Within this set of realizations
violent fluctuations of complexity related observables are observed.
The search time also shows
an exponential behavior of the form $\tau_s^0/N^2={\rm a}+{\rm b}~{\rm exp}[cB_0]$
with $c\approx 0.8$.
Thus the low-temperature free energy landscape of the 3-SAT problem
has the simple property that the static free energy barrier $B_0$
determines the ground-state search dynamics.

We now quantize the problem by introducing
a standard Trotter-Suzuki time discretization~\cite{Suzuki}.
We choose a regular temporal lattice with $N_\tau=128,256$
time-slices, a finite step-size $\Delta \tau=1$ in $\tau$ direction and
periodic boundary conditions in Trotter time. The inverse temperature
is $\beta = N_\tau \Delta \tau$ and the Boltzmann
factor of the quantized problem at imaginary time is
\begin{eqnarray}
\ln [P_{\rm B,q}]= - \kappa_0 \sum_{\tau=1}^{N_\tau} H_0(\lbrace s_1(\tau),...,s_N(\tau) \rbrace) \nonumber \\
-\kappa_\tau \sum_i^N \sum_{\tau=1}^{N_\tau}   s_i(\tau) s_i(\tau+1),
\end{eqnarray}
with positive ferromagnetic hopping parameters
$\kappa_0 =  \lambda \Delta \tau$ and $\kappa_\tau =  - \ln [{\rm tanh}((1-\lambda)\Delta \tau)]/2$ \cite{Suzuki}.
These equations implement
the AQC partition function Eq.~(\ref{partition_function})
as a function of $\lambda$ and $\beta$, up to discretization
errors caused by the finiteness of the
regularization. The first energy gap $\xi^{-1}$
is obtained from a large $\tau$ fit to the exponential decay of the expectation value
of the connected two point function in the canonical mean
\be
\Gamma(\tau)=\langle {\cal O}(0) {\cal O}(\tau)\rangle - \langle {\cal O}(0)\rangle ^2 ~\propto e^{-{\tau\over\xi}}~.
\label{quantum_correlator}
\ee
We have experimented with several time-local observables
${\cal O}(\tau)$ and found that among an extended set of trial observables
the $\tau$-dependent ground-state overlap
$ {\cal O}{\rm gs}(\tau)= N^{-1}\sum_{i=1}^N s_i(\tau) s_i^{\rm gs} $
yields the best statistical signals for the exponential
decay with $\tau$.
The minimum energy gap, or the maximum spin-spin
correlation length  $\xi_{\rm max}$
determines the complexity. We find it by a search,
that uses parallel tempering simulations in the control parameter $\lambda$ \cite{PATE}
on a $\lambda$ partition with $64$ elements. Elementary low-temperature exchange updates
are absolutely essential for error reduction in the quantum correlator.

\begin{figure}[t]
\centering
\psfrag{LABEL1}[t][t][2][0]{$\lambda$}
\psfrag{LABEL2}[t][t][2][0]{$\langle o_{\rm gs}\rangle$}
\psfrag{LABEL3}[t][t][2][0]{$\tau$}
\psfrag{LABEL4}[t][t][2][0]{$\ln~\Gamma(\tau)$}
\psfrag{L5}[t][t][2][0]{$\lambda^*$}
\includegraphics[angle=-90,width=0.55\textwidth]{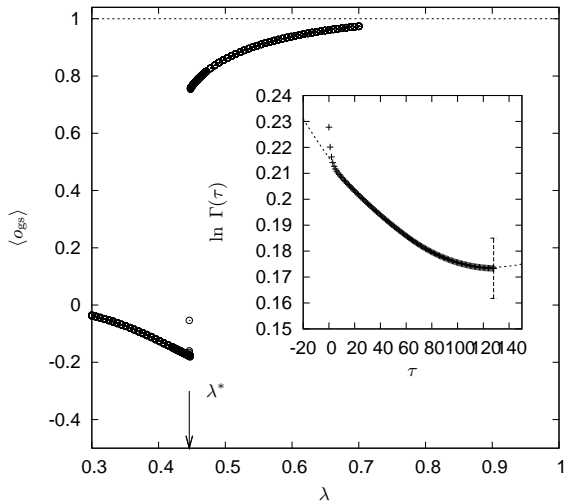}
\caption{Quantum limit: Signature of the quantum phase transition
for the same realization of Fig.~\ref{fig:fig1}.
The mean overlap observable $\langle o_{\rm gs}\rangle$
as a function of $\lambda$
shows a jump at $\lambda^*\approx 0.446$.
Inset: Overlap-overlap correlation
function $\Gamma(\tau)$ at $\lambda^*$ as a function of imaginary time $\tau$.
\label{fig:fig3}}
\end{figure}

We have studied the quantum phase transition of the partition function
Eq.~(\ref{partition_function}) for $r=8$
and $N$ values up to $N=80$. In Fig.~\ref{fig:fig3} we display the expectation value of the
quantum ground-state overlap observable
$\langle o_{\rm gs}\rangle = N_\tau^{-1} \sum_{N_\tau}o_{\rm gs} (\tau)$ for a specific $N=80$
realization. A quantum phase
transition,  which is of blatant discontinuous
nature, is observed at $\lambda^*$.
Interestingly, the transition
to the ground-state proceeds with increasing $\lambda$
from configurations that are slightly anti-parallel
to the known ground state, see the negative values of $o_{\rm gs}$
for $\lambda < \lambda^{*}$.
The inset of Fig.~\ref{fig:fig3}
displays the quantum correlation
function $\Gamma(\tau)$ of Eq.~(\ref{quantum_correlator})
at $\lambda^*$, which for the specific example
decays exponentially at the numerical value
$\xi_{\rm max}\approx 185$.

\begin{figure}[t]
\centering
\psfrag{LABEL1}[t][t][2][0]{$B_0$}
\psfrag{LABEL2}[t][t][2][0]{$\ln \xi$}
\psfrag{LABEL3}[t][t][2][0]{$N$}
\psfrag{LABEL4}[t][t][2][0]{$\ln \langle\xi\rangle$}
\psfrag{SOMETEXT}[t][t][2][0]{$\ln \langle\xi\rangle$}
\includegraphics[angle=-90,width=0.55\textwidth]{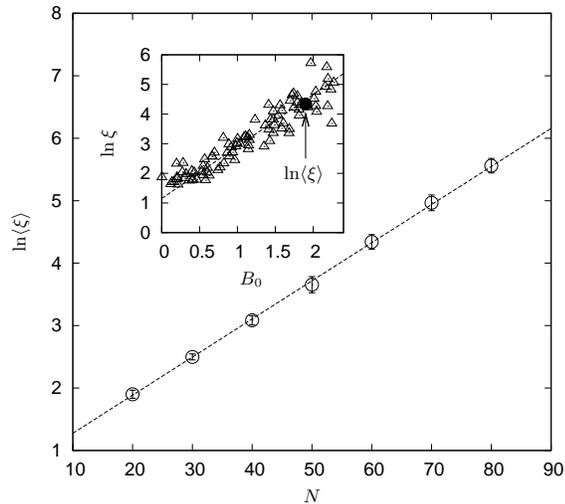}
\caption{Quantum limit: Finite-size scaling of $\ln \langle\xi\rangle$ as a function of $N$ for $r=8$.
The largest median correlation length is $\langle\xi\rangle =259.9$ at $N=80$.
The straight line displays the fit Eq.~(\ref{fit_to_xi}). Inset:
Scatter plot
of tupels of $\ln \xi$ and $B_0$ (triangles) for $N=60$.
The data exhibit a linear correlation (straight line).
We determine the correlation length at the position of the median
barrier $\langle B_0\rangle =1.90(6)$ (arrow and solid circle).
\label{fig:fig4}}
\end{figure}

In the analysis of the quantum complexity we have picked a
subset of $91$ realizations for $N=60$ at about median - or less - classical
complexity given by the $B_0$ value.
In the inset of Fig.~\ref{fig:fig4} we display the correlation of $B_0$ values
with their quantum counterparts, the maximal
spin-spin correlation length $\xi_{\rm max}$ values.
We observe a linear correlation between $\ln \xi_{\rm max}$ and $B_0$,
see the straight line, and not even the slightest
tendency that the quantum complexity
is weaker than the corresponding classical one.
A similar observation is made for
all $N$.

We simplify the calculation of the quantum complexity by use of the median.
Note that median averages do not
necessarily require the actual calculation
of extremal values, but just of those in the vicinity of the median. Given our data, we may
determine the spin-spin correlation length $\xi$
at the median position of the nucleation barrier $\langle B_0\rangle$.
For correlated data the latter quantity is a
good estimator of the spin-spin
correlation length in the median
$\langle\xi\rangle$. A fit to the
form
\be
\langle\xi\rangle~=~A~e^{g_qN},
\label{fit_to_xi}
\ee
on the entire $N$-interval $20 \le N \le 80$ with
$\chi_{\rm d.o.f}=0.1$ yields for the quantum growth rate constant $g_q=0.061(1)$
for the 3-SAT problem with $r=8$.
Figure~\ref{fig:fig4} shows
$\ln \langle \xi \rangle$ as a function of $N$ (circles) and the
fit in Eq.~(\ref{fit_to_xi}) (straight line).
As can be clearly seen, the data are not compatible with a polynomial behavior.
Following the Landau-Zener theory, twice the growth rate constant $2 g_q$ should be considered
for the comparison to the classical search time complexity.

In summary, we have determined the exponential
singularities that dominate
the classical and quantum
running times for ground-state
searches in the median
of an ensemble
in the 3-SAT problem with unique satisfiability assignments.
For $r=8$
we measure the growth rate constants
$g_c=0.016(1)$ and
$g_q=0.061(2)$
of the corresponding classical $\tau_s^0$ and
quantum $\xi_{max}$ exponential behavior in the ground-state searches.
Our numerical data are precise and
the classical growth rate constant is confirmed
by a static free energy i.e., nucleation barrier $B_0$ scaling.
The finite size scaling window is large
and excludes a polynomial
behavior for quantum and classical searches.
For the case of standard AQC we find that twice the growth rate
constant $2g_q$ undershoots the Grover value $g=\ln 2/2$.
Hence, standard AQCs constitute a
class of ground-state searches, that
compares favorable to Grovers's
quantum search.
However, there is no indication that for
the 3-SAT problem standard
AQCs can reduce the
exponential complexity to a polynomial one. Thus, standard
AQC ground-state searches for NP-hard problems are not quite as
powerful \cite{VanDam_2001} as was conjectured
earlier \cite{Farhi}. We also find a quantum growth rate constant
that exceeds the classical one by a factor
of 3.8. Therefore, sole quantum
fluctuations are even significantly less efficient
than sole classical fluctuations.
In the future one may determine complexities
of systems with modified driver Hamiltonians, that
are optimized with respect to ground state searches.
An alternative quantitative approach to computational intractability
can rely on free energy calculations for
Instantons \cite{Coleman}.

T.N. thanks the theory department
at Bielefeld university
for extended hospitality. Calculations
were performed on the
JUMP and JUROPA supercomputers at JSC
and on the NICOLE workstation cluster
of NIC (VSR grant JJSC02).
This work is partially supported
by NCF, the Netherlands.

\end{document}